\documentclass[%
,twocolumn%
,secnumarabic%
,amssymb,aps,pra,nobibnotes]{revtex4-1}
\usepackage{docs}
\usepackage{float}
\usepackage{graphicx}
\usepackage{dcolumn}
\usepackage{bm}
\usepackage{appendix}

\usepackage[dvipdfm,  
            pdfstartview=FitH,
            CJKbookmarks=true,
            bookmarksnumbered=true,
            bookmarksopen=true,
            linkcolor=green,
            anchorcolor=green,
            citecolor=green
            ]{hyperref}

\expandafter\ifx\csname package@font\endcsname\relax\else
 \expandafter\expandafter
 \expandafter\usepackage
 \expandafter\expandafter
 \expandafter{\csname package@font\endcsname}%
\fi
\DeclareRobustCommand\substyle{\name@idx{document substyle}}%
\DeclareRobustCommand\classoption{\name@idx{document class option}}%
\DeclareRobustCommand\classname{\name@idx{document class}}%
\def\name@idx#1#2{%
 {\ttfamily#2}%
 \index{#2\space#1=\string\ttt{#2}\space#1}\index{#1>#2=\string\ttt{#2}}%
}%

\begin{document}
\title{Finite-size analysis of continuous-variable measurement-device-independent \\quantum key distribution}%
\author{Xueying Zhang$^1$}
\author{Yichen Zhang$^{1,}$}
\email{zhangyc@bupt.edu.cn}
\author{Yijia Zhao$^1$}
\author{Xiangyu Wang$^1$}
\author{Song Yu$^{1,}$}
\email{yusong@bupt.edu.cn}
\author{Hong Guo$^{1,2}$}

\affiliation{$^1$State Key Laboratory of Information Photonics and Optical Communications, Beijing University of Posts and Telecommunications, Beijing 100876, China}

\affiliation{$^2$State Key Laboratory of Advanced Optical Communication Systems and Networks, School of Electronics Engineering and Computer Science, Center for Quantum Information Technology, Center for Computational Science and Engineering, Peking University, Beijing 100871, China}

\date{\today}%


\begin{abstract}
We study the impact of the finite-size effect on the continuous-variable measurement-device-independent quantum key distribution (CV-MDI QKD) protocol, mainly considering the finite-size effect on the parameter estimation procedure. The central-limit theorem and maximum likelihood estimation theorem are used to estimate the parameters. We also analyze the relationship between the number of exchanged signals and the optimal modulation variance in the protocol. It is proved that when Charlie's position is close to Bob, the CV-MDI QKD protocol has the farthest transmission distance in the finite-size scenario. Finally, we discuss the impact of finite-size effects related to the practical detection in the CV-MDI QKD protocol. The overall results indicate that the finite-size effect has a great influence on the secret key rate of the CV-MDI QKD protocol and should not be ignored.

\end{abstract}
\maketitle

\section{INTRODUCTION}
Quantum key distribution (QKD) \cite{bennett1984g,ekert1991quantum,RevModPhys.74.145.2002,scarani2009security} is the most mature technology in quantum cryptography, which allows two distant legitimate parties, Alice and Bob, to generate secret keys through an untrusted channel controlled by an eavesdropper, Eve. The coherent state continuous-variable QKD protocol (CV-QKD) \cite{RevModPhys5132005,RevModPhys6212012}, based on continuous modulation (Gaussian modulation), was proposed in 2002 \cite{grosshans2002continuous}. To effectively improve protocol security transmission distance, a continuous-variable two-way quantum key distribution protocol was proposed \cite{jouguet2008Continuous,Ottaviani2015Two-way}. After that, the CV-QKD protocol has received increasing attention in the past few years due to its high secret key rate and low-cost advantage \cite{Weedbrook2004,grosshans2005collectiveattacks,Navascues2005,Wolf2006,Navascues2006,Garcia2006}.  In the experiment, Jouguet \emph{et al.}\cite{jouguet2013experimental} achieved the CV-QKD experiment of all-fiber Gaussian modulation coherent state and homodyne detection of 80 km in the laboratory. The field tests of a CV-QKD system \cite{Zhang2017commercial} have extended the distribution distance to 50 km over commercial fiber, where the secure key rates are two orders of magnitude higher than previous field test demonstrations.

Although the QKD protocols, including CV protocols \cite{Garcia2006,Navascues2006,Muller-Quade2009Composability}, are guaranteed to have unconditional security in theory based on quantum physics, the actual security is related to the performance of the device. Due to the imperfection of the detector in the practical CV-QKD system, Eve can implement quantum hacking attacks for CV detectors, such as the local oscillator calibration attack \cite{ma2013local,jouguet2013preventing}, the wavelength attack \cite{huang2013quantum, ma2013wavelength}, the detector saturation attack \cite{qin2016saturation}. These attacks use the detector's imperfect characteristics to operate the detector results to reduce the additional noise variance, so that Alice and Bob estimate the secret key rate too high, resulting in security risks. For this reason, continuous-variable measurement-device-independent (CV-MDI) protocol has been proposed \cite{li2014continuous, pirandola2015high}. Since the measurement part of the protocol is completely committed by an untrusted third party, the security of the protocol does not depend on the security of the detector. Therefore, this protocol can naturally resist all hacker attacks against detectors.

In most conventional security proofs, the theoretical security analysis relies on the assumption that the two communication parties exchange the infinite number of signals in the asymptotic scheme \cite{scarani2009security}. Hence, the finite-size effect is the key problem that CV protocol needs to be solved in the practical implementation. In recent years, the attention of researchers has gradually shifted to the practical security of CV-QKD protocol. Scarani and Renner \cite{scarani2008quantum} proposed a security bound based on smooth min-entropy in 2008. In 2010, Leverrier \emph{et al.} \cite{leverrier2010finite} extended the finite-size analysis framework from discrete-variable (DV) protocol to CV protocol, taking into account the finite-size effect of the parameter estimation process. In 2012, Jouguet \emph{et al.} \cite{jouguet2012analysis} researched the finite-size effect of CV-QKD protocol on the secret key rate in the case of practical detections. So far, the security of the CV-MDI QKD protocol in the asymptotic scenario has been demonstrated \cite{li2014continuous,pirandola2015high,zhang2014continuous,PhysRevA.91.022320,Zhang2015Noiseless}. But the feasibility of CV-MDI QKD protocol under finite-size effect has not yet been confirmed.

To solve this problem, we study the influence of the finite-size effect on the CV-MDI QKD protocol under collective attack. Here we only consider the reverse reconciliation protocol and for the direct reconciliation protocol we can use the similar calculation method. The numerical simulations of the protocol are given at block lengths between ${10^6}$ and ${10^{10}}$. When Charlie is placed in Bob (asymmetric case) and considering the optimal modulation variance conditions, the farthest transmission distance can reach 86km for reconciliation efficiency $\beta  = 1$ and 75 km for $\beta  = 96.9\% $. Finally, we discuss the impact of the practical detection on the CV-MDI QKD protocol under the finite-size effect.

The rest of this paper is organized as follows: In Sec. II, we first introduce the main idea of the CV-MDI QKD protocol, and then study the finite-size analysis of the parameter estimation process of CV-MDI QKD protocol in detail, and introduce the method and formulas that are used in finite-size analysis. In Sec. III, we show the simulation results of the secret key rate, and give the optimal modulation variance and the maximal transmission distance under different conditions. Our conclusions are drawn in Sec. IV.

\begin{figure}[t]
\centering\includegraphics[width=3.5in]{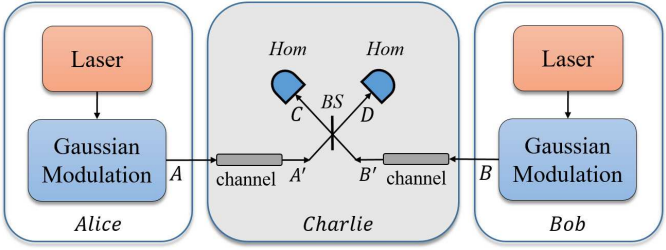}
\caption{\label{fig:epsart}A schematic diagram of CV-MDI QKD. BS stands for 50 : 50 beam splitter, Hom stands for homodyne detection.} 
\label{fig:setup}
\end{figure}

\section{CV-MDI QKD PROTOCOL AND PARAMETER ESTIMATION}

\begin{figure}[htbp]
\centering\includegraphics[width=3.0in]{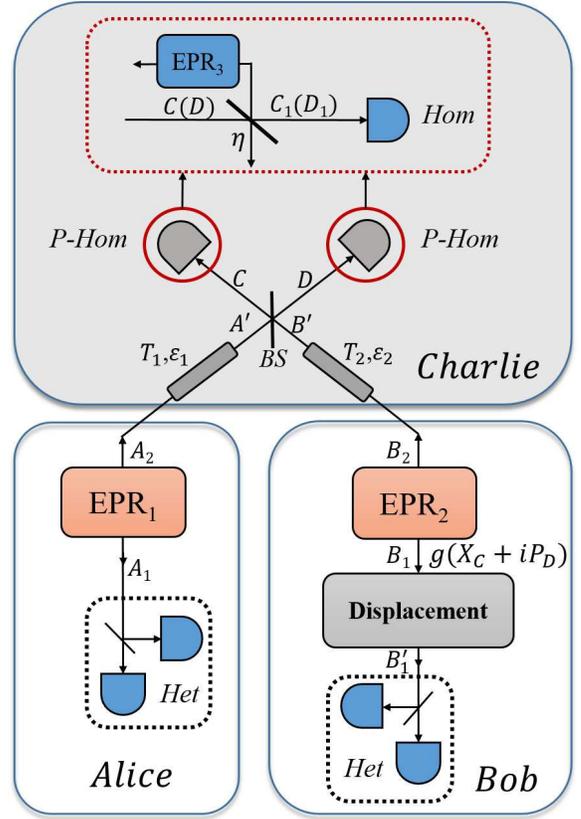}
\caption{\label{fig:epsart}Entanglement-based scheme of CV-MDI QKD with practical detector. ${T_1}({T_2})$ is the channel transmittance for Alice-Charlie (Bob-Charlie), ${\varepsilon _1}({\varepsilon _2})$ is the channel excess noise for Alice-Charlie (Bob-Charlie). $\eta $ is the efficiency of the detection, $v$ is the variance of the thermal state, ${v_{el}}$ is the variance of electronic noise. EPR is the two-mode squeezed state, P-Hom is practical homodyne detection, Hom is ideal homodyne detection, Het is heterodyne detection.}
\label{fig:EB}
\end{figure}

The setup of the CV-MDI QKD protocol is illustrated in Fig. \ref{fig:setup}. The Gaussian modulation coherent state CV-MDI QKD protocol is mainly described as follows: First, Alice and Bob use lasers to generate light sources, and prepare Gaussian coherent states independently by using the phase modulator and amplitude modulator, respectively. Then, Alice and Bob send the prepared quantum states to an untrusted third party (Charlie) for continuous-variable Bell measurement. The specific operation is that Charlie receives two quantum states through a 50 : 50 beam splitter, and then uses two homodyne detectors to measure the interference results. Next, Charlie publishes his measurement results. Finally, Alice and Bob use the corrected data for parameter estimation, data postprocessing, data reconciliation, and private key amplification to get the final security secret key.

The above is the protocol process of the CV-MDI QKD protocol in the prepare-and-measure (PM) scheme. The PM scheme is easy to implement, but it often uses the entanglement-based (EB) scheme in security analysis. The equivalence between the EB scheme of the CV-MDI QKD protocol and the PM scheme has been proven in Ref. \cite{li2014continuous}, and the EB scheme of the CV-MDI QKD protocol is shown in Fig. \ref{fig:EB}. The equivalence is reflected in two points: On the one hand, the equivalence of the state preparation; that is, the heterodyne detection of one mode of the two-mode squeezed (EPR) state is equivalent to the preparation of a Gaussian modulation coherent state. On the other hand, the data correction process in the PM scheme is equivalent to the displacement operation in the EB scheme. Therefore, only the EB version of the CV-MDI QKD protocol is analyzed in the following security analysis.

To calculate the secret key rate of the CV-MDI QKD protocol, the researchers proposed two calculation methods \cite{li2014continuous,pirandola2015high}: One is the secret key rate formula based on entanglement swapping. That is, assuming that Bob's EPR state preparation and displacement operations are regarded as manipulated by Eve in Fig. \ref{fig:EB}, then the CV-MDI QKD protocol can be equivalent to a one-way CV-QKD protocol. The other is the secret key rate formula based on the sub-scenario. That is, each subscenario is the same as a one-way CV-QKD protocol, and the final secret key rate is the mean of the key rate of each sub-scenario. CV-MDI QKD protocol in the asymptotic regime and collective attack conditions, the secret key rate of the former is smaller than the latter's secret key rate. Moreover, in the case of finite-size, the two security analysis methods are equivalent to the coherent state CV-QKD one-way protocol using heterodyne detection. In the following, because the idea of security analysis is more concise and the calculation of the secret key rate is simpler, we adopt the former secret key rate calculation method.

According to the finite-size effect analysis of discrete-variable QKD protocol against collective attacks, the expression of the secret key rate of the one-way CV-QKD protocol against collective attack in the finite-size case is \cite{leverrier2010finite}:
\begin{equation}
k=\frac{n}{N}[\beta I(a:b)-{S_{{\epsilon _{_{PE}}}}}(b:E)-\Delta (n)],
\label{eq:finiteK}
\end{equation}
where $N$ is the total number of signals exchanged by Alice and Bob, of which only $n$ signals are used to generate the keys. $\beta  \in [0,1]$ is the reconciliation efficiency, and $I(a:b)$ is the mutual information of Alice and Bob. Considering the influence of the finite-size effect on the accuracy of the parameter estimation; that is, under certain failure probability $\epsilon_{PE}$, the true channel parameters are within a certain confidence interval near the estimated parameters, then the conditional entropy of Eve and Bob is expressed as ${S}_{\epsilon_{PE}}(b:E)$. The most important parameter in the expression, $\Delta (n)$, is related to the security of the private key amplification \cite{leverrier2010finite}. Its value is given by:
\begin{equation}
\Delta (n)=(2dim{H}_{X}+3)\sqrt{\frac{{log}_{2}(2/\tilde{\epsilon })}{n}}+\frac{2}{n}{log}_{2}(1/{\epsilon _{PA}}).
\label{eq:finite-n}
\end{equation}

In Eq. (\ref{eq:finiteK}), $\Delta (n)$ is the correction term in the formula for security secret key rate, which varies with the total length of the exchanged signals. The first term of Eq. (\ref{eq:finite-n}) is the convergence speed of the smooth minimum entropy of an independent and identically distributed state to the von Neumann entropy, which is the main part to determine $\Delta (n)$. Where $H_X$ corresponds to the dimension of the emph{Hilbert} space of the variable $x$ in the raw key, taking $\dim {H_X} = 2$ \cite{leverrier2010finite} in the CV protocol. $\bar \epsilon $ and $\epsilon_{PA}$ are the smoothing parameter and the failure probability of private key amplification processes, respectively, and we take their optimal value as $\bar \epsilon  = {\epsilon _{PA}} = {10^{ - 10}}$ \cite{leverrier2010finite}.

In the following, we study the impact of the finite-size effect on the parameter estimation process of the CV-MDI QKD protocol. More precisely, we analyze the influence of finite signals length $N$ on the estimation of channel excess noise $ {\varepsilon _1},{\varepsilon _2}$. Due to the limited signals length, the statistical fluctuation of the sampling estimation in the parameter estimation process will be worse, which makes the evaluation accuracy of both communication sides on Eve's eavesdropping behavior worse. To ensure the security of the protocol, it is necessary to do the worst estimate of the impact of eavesdropping. That is, one need to compute the maximum value of the Holevo information between Eve and Bob in the case of statistical fluctuation in the parameter estimation, the maximum value of ${S_{{\epsilon _{PE}}}}(b:E)$.

The calculation of ${S}_{ {\epsilon}_{PE}}(b:E)$ depends on the covariance matrix ${\gamma _{{A_1}{{B'}_1}}}$ of  shared state ${\rho _{{A_1}{{B'}_1}}}$ of Alice and Bob in the EB version of the protocol. In particular, in the CV-MDI QKD protocol, we need to consider the statistical fluctuation of the channel transmittance (${T}_{1}$ ,${T}_{2}$) and the excess noise of the Alice-Charlie channel and the Bob-Charlie channel (${\varepsilon }_{1}$, ${\varepsilon }_{2}$), the Alice modulation variance (${V}_{A}$), and the Bob modulation variance (${V}_{B}$), respectively.

In the EB version of the CV-MDI QKD protocol, as shown in Fig. \ref{fig:setup}, before Charlie makes a homodyne detection to the $C$ and $D$ modes respectively, ${\gamma _{{A_1}CD{B_1}}}$ takes the following form:
\begin{widetext}
\begin{equation}
{\gamma_{\!{A_1}\!C\!D\!{B_1}}}\! =\!\left(\! {\begin{array}{*{20}{c}}
{{V_1}{I_2}}&{\sqrt {\frac{1}{2}{T_1}\left( {V_1^2\! -\! 1} \right)} {\sigma _z}}&{\sqrt {\frac{1}{2}{T_1}\left( {V_1^2\! -\! 1} \right)} {\sigma _z}}&{0{I_2}}\\
\!{\sqrt {\frac{1}{2}{T_1}\left( {V_1^2\! -\! 1} \right)} {\sigma _z}}&{\left( {\frac{1}{2}{T_1}\left( {{V_1} + {\chi _1}} \right)\! + \!\frac{1}{2}{T_2}\left( {{V_1} \!+ \!{\chi _2}} \right)} \right)\!{I_2}}&{\left( {\frac{1}{2}{T_1}\left( {{V_1}\! +\! {\chi _1}} \right)\! - \!\frac{1}{2}{T_2}\left( {{V_1}\! + \!{\chi _2}} \right)} \right)\!{I_2}}&{\sqrt {\frac{1}{2}{T_2}\left( {V_2^2 - 1} \right)} {\sigma _z}}\\
\!{\sqrt {\frac{1}{2}{T_1}\left( {V_1^2\! -\! 1} \right)} {\sigma _z}}&{\left( {\frac{1}{2}{T_1}\left( {{V_1}\! +\! {\chi _1}} \right)\! -\! \frac{1}{2}{T_2}\left( {{V_1}\! +\! {\chi _2}} \right)} \right)\!{I_2}}&{\left( {\frac{1}{2}{T_1}\left( {{V_1}\! +\! {\chi _1}} \right)\! + \!\frac{1}{2}{T_2}\left( {{V_1}\! +\! {\chi _2}} \right)} \right)\!{I_2}}&{\! -\! \sqrt {\frac{1}{2}{T_2}\left( {V_2^2 - 1} \right)} {\sigma _z}}\\
\!{0{I_2}}&{\sqrt {\frac{1}{2}{T_2}\left( {V_2^2\! -\! 1} \right)} {\sigma _z}}&{ \!-\! \sqrt {\frac{1}{2}{T_2}\left( {V_2^2\! -\! 1} \right)} {\sigma _z}}&{{V_2}{I_2}}
\end{array}} \!\right)\!,
\label{gama-wide}
\end{equation}
\end{widetext}
where ${I_2}$ is the $2 \times 2$ identity matrix and ${\sigma _z} = \left[ {\begin{array}{*{20}{c}}
1&0\\
0&{ - 1}
\end{array}} \right]$, so ${\gamma _{{A_1}CD{B_1}}}$ is obviously an $8 \times 8$ covariance matrix. The parameters of the above matrix are given by ${V}_{1}={V}_{A}+1,{V}_{2}={V}_{B}+1$, ${V}_{A}$ and ${V}_{B}$ are the modulation variance of Alice and Bob in the PM protocol. ${T}_{1}$ and ${T}_{2}$ are the transmittance of the Alice-Charlie channel and the Bob-Charlie channel. ${\varepsilon }_{1}$ and ${\varepsilon }_{2}$ in ${\chi }_{1}=\frac{1}{{T}_{1}}-1+{\varepsilon }_{1}$, ${\chi }_{2}=\frac{1}{{T}_{2}}-1+{\varepsilon }_{2}$ are the excess noise of the corresponding channel. According to the covariance matrix Eq. (\ref{gama-wide}), we know that the modulation variance of Alice and Bob, $\left\langle {x_1^2} \right\rangle $and $\left\langle {x_2^2} \right\rangle $, the variance of Charlie, $\left\langle {y_1^2} \right\rangle $, $\left\langle {y_2^2} \right\rangle $ and $\left\langle {{y_1}{y_2}} \right\rangle $, the covariance of Alice and Charlie $\left\langle {{x_1}{y_1}} \right\rangle $, and the covariance of Bob and Charlie $\left\langle {{x_2}{y_2}} \right\rangle $. These values and secret key rate parameters are related through:
\begin{equation}
\left\langle {x_1^2} \right\rangle  = {V_1} - 1 = {V_A},\nonumber\\
\left\langle {x_2^2} \right\rangle  = {V_2} - 1 = {V_B},
\end{equation}
\begin{equation}
\left\langle {y_1^2} \right\rangle {\rm{ \nonumber\\
= }}\left\langle {y_2^2} \right\rangle  \nonumber\\
= \frac{1}{2}({T_1}{V_A} + {T_2}{V_B}) + 1 + \frac{1}{2}({T_1}{\varepsilon _1} + {T_2}{\varepsilon _2}),
\end{equation}
\begin{eqnarray}
\left\langle {{x_1}{y_1}} \right\rangle  = \sqrt {\frac{{{T_1}}}{2}} ({V_1} - 1) = \sqrt {\frac{{{T_1}}}{2}} {V_A},\nonumber\\
\left\langle {{x_2}{y_2}} \right\rangle  = \sqrt {\frac{{{T_2}}}{2}} ({V_2} - 1) = \sqrt {\frac{{{T_2}}}{2}} {V_B},
\end{eqnarray}
\begin{equation}
\left\langle {{y_1}{y_2}} \right\rangle  = \frac{1}{2}({T_1}{V_A} - {T_2}{V_B}) + \frac{1}{2}({T_1}{\varepsilon _1} - {T_2}{\varepsilon _2}).
\end{equation}

In practical CV-MDI QKD systems, the quantities estimated by Alice and Bob are obtained by the sampling of $m=N-n$ pairs of correlated variables ${({x}_{i},{y}_{i})}_{i=1\ldots m}$. Since the Alice-Charlie channel and the Bob-Charlie channel are linear channels, the variables of Alice, Bob and Charlie follow a Gaussian distribution. Within this model, we consider that the data before the beam splitter is represented as ${y}_{1}'$, ${y}_{2}'$. So before the beam splitter, Alice and Charlie's and Bob and Charlie's data are linked through the following relation:
\begin{eqnarray}
&&{y}_{1}'={t}_{1}'{x}_{1}+{z}_{1},\nonumber\\
&&{y}_{2}'={t}_{2}'{x}_{2}+{z}_{2},
\end{eqnarray}
where ${t}_{1}'=\sqrt{{T}_{1}},{t}_{2}'=\sqrt{{T}_{2}}$, ${z}_{1},{z}_{2}$ follow a centered normal distribution with unknown variance ${{\sigma }_{1}'}^{2}=1+{T}_{1}{\varepsilon }_{1},{{\sigma }_{2}'}^{2}=1+{T}_{2}{\varepsilon }_{2}$. According to the known parameter relations in the covariance matrix, it can be expressed that the variance of the unknown parameters before the beam splitter:
\begin{eqnarray}
\langle {{y'_1}^2}\rangle &&=\left\langle {{y_1}^2} \right\rangle  + \left\langle {{y_1}{y_2}} \right\rangle  \nonumber\\
&&= {T_1}{V_A} + 1 + {T_1}{\varepsilon _1} = {t'_1}^2{V_A} + {\sigma '_{1}}^{2},
\end{eqnarray}
\begin{eqnarray}
\langle {{y'_2}^2} \rangle  &&= \left\langle {{y_2}^2} \right\rangle  - \left\langle {{y_1}{y_2}} \right\rangle \nonumber\\
&&= {T_2}{V_B} + 1 + {T_2}{\varepsilon _2} = {t'_2}^2{V_B} + {\sigma '_{2}}^{2}.
\end{eqnarray}
Maximum-likelihood estimators ${\hat{t}}_{1}',{\hat{t}}_{2}',{\hat{\sigma'_{1}}}^{2},{\hat{\sigma'_{2}}}^{2},{\hat{V}}_{A},{\hat{V}}_{B}$ are known for the normal linear model \cite{leverrier2010finite}:
\begin{eqnarray}
&&\hat{t}'_{1}=\frac{\sum_{i=1}^{m}{x}_{1i}{y}_{1i}'}{\sum_{i=1}^{m}{{x}_{1i}}^{2}},\nonumber\\
&&\hat{t}'_{2}=\frac{\sum_{i=1}^{m}{x}_{2i}{y}_{2i}'}{\sum_{i=1}^{m}{{x}_{2i}}^{2}},
\end{eqnarray}
\begin{eqnarray}
&&{\hat{\sigma'_{1}}}^{2}=\frac{1}{m}\sum_{i=1}^{m}{\left({y}'_{1i}-{\hat{t}'}_{1}{x}_{1i} \right)}^{2},\nonumber\\
&&{\hat{\sigma'_{2}}}^{2}=\frac{1}{m}\sum_{i=1}^{m}{\left({y}'_{2i}-{\hat{t}'}_{2}{x}_{2i} \right)}^{2},
\end{eqnarray}
\begin{eqnarray}
{\hat V_A}=\frac{1}{m}\sum_{i=1}^{m}{{x}_{2i}}^{2},\nonumber\\
{\hat V_B}=\frac{1}{m}\sum_{i=1}^{m}{{x}_{1i}}^{2}.
\end{eqnarray}
The estimators ${{\hat t'}_1},{{\hat t'}_2},{\hat{\sigma'_{1}}}^{2},{\hat{\sigma'_{2}}}^{2},{\hat V_A},{\hat V_B}$ are independent estimators with the following distributions:
\begin{equation}
{{\hat t'}_1}\sim{\rm N}\left( {{t_1^{'}},\frac{{\sigma '_{1}}^{2}}{{\sum\nolimits_{i = 1}^m {x_{1i}^2} }}} \right),{{\hat t'}_2}\sim{\rm N}\left( {{t^{'}_2},\frac{{\sigma '_{2}}^{2}}{{\sum\nolimits_{i = 1}^m {x_{2i}^2} }}} \right)
\end{equation}
\begin{equation}
\frac{m{\hat{\sigma'_{1}}}^{2}}{{\sigma '_{1}}^{2}},\frac{m{\hat{\sigma'_{2}}}^{2}}{{\sigma '_{2}}^{2}},\frac{{m{{\hat V}_A}}}{{{V_A}}},\frac{{m{{\hat V}_B}}}{{{V_B}}}\sim{\chi ^2}(m - 1),
\end{equation}
where ${t'_1},{t'_2},{\sigma '_{1}}^{2},{\sigma '_{2}}^{2},{V_A},{V_B}$ are the true values of the parameters. Due to the limit of length $m$, we can estimate the confidence interval for these parameters when the confidence probability is ${\epsilon _{PE/2}}$:
\begin{eqnarray}
&&{t^{'}_1} \in [{{\hat t'}_1} - \Delta {{t'}_1},{{\hat t'}_1} + \Delta {{t'}_1}],\nonumber\\
&&{t^{'}_2} \in [{{\hat t'}_2} - \Delta {{t'}_2},{{\hat t'}_2} + \Delta {{t'}_2}],
\end{eqnarray}
\begin{eqnarray}
&&{\sigma '_{1}}^{2} \in [{\hat{\sigma'_{1}}}^{2} - \Delta {\sigma '_{1}}^{2},{\hat{\sigma'_{1}}}^{2} + \Delta {\sigma '_{1}}^{2}],\nonumber\\
&&{\sigma '_{2}}^{2} \in [{\hat{\sigma'_{2}}}^{2} - \Delta {\sigma '_{2}}^{2},{\hat{\sigma'_{2}}}^{2} + \Delta {\sigma '_{2}}^{2}],
\end{eqnarray}
\begin{eqnarray}
&&{V_A} \in [{{\hat V}_A} - \Delta {V_A},{{\hat V}_A} + \Delta {V_A}],\nonumber\\
&&{V_B} \in [{{\hat V}_B} - \Delta {V_B},{{\hat V}_B} + \Delta {V_B}],
\end{eqnarray}
where $\Delta {t'_1} = {z_{{\epsilon _{PE}}/2}}\sqrt {\frac{{\hat{\sigma'_{1}}}^{2}}{{m{V_A}}}},~\Delta {t'_2} = {z_{{\epsilon _{PE}}/2}}\sqrt {\frac{{\hat{\sigma'_{2}}}^{2}}{{m{V_A}}}},\\\Delta {\sigma '_{1}}^{2} = {z_{{\epsilon _{PE}}/2}}\frac{{\hat{\sigma'_{1}}}^{2}\sqrt 2}{{\sqrt m }},~\Delta {V_A} = {z_{{\epsilon _{PE}}/2}}\frac{{{{\hat V}_A}\sqrt 2 }}{{\sqrt m }},\\\Delta {\sigma '_{2}}^{2} = {z_{{\epsilon _{PE}}/2}}\frac{{\hat{\sigma'_{2}}}^{2}\sqrt 2 }{{\sqrt m }},\Delta {V_B} = {z_{{\varepsilon _{PE}}/2}}\frac{{{{\hat V}_B}\sqrt 2 }}{{\sqrt m }}$, and ${z_{{\epsilon _{PE}}/2}}$ satisfies $1 - erf({z_{{\epsilon _{PE}}/2}}/\sqrt 2 )/2 = {\epsilon _{PE}}/2$. ${\epsilon _{PE}}$ is the failure probability of the parameter estimation process, which generally takes $10^{-10}$.~$erf(x)$ is error function, defined as
\begin{equation}
erf(x) = \frac{2}{{\sqrt \pi  }}\int_0^x {{e^{ - {t^2}}}} dt.
\end{equation}
We can estimate ${T_1} ={\hat{t'_{1}}}^{2},{T_2} = {\hat{t'_{2}}}^{2}$ and ${\varepsilon _1} = \frac{{{\hat{\sigma'_{1}}}^{2} - 1}}{{\hat{t'_{1}}}^{2}},{\varepsilon _2} = \frac{{\hat{\sigma'_{2}}}^{2}-1}{{\hat{t'_{2}}}^{2}}$ by previous estimators and confidence intervals.
Next, Charlie uses two homodyne detectors to measure the C and D modes, and Bob performs displacement operations based on Charlie's measurements. Indeed, the CV-MDI QKD protocol is equivalent to a one-way CV-QKD protocol when Bob's EPR state preparations and displacement operations are also untrusted. Then the covariance matrix ${\gamma _{{A_1}{B_{1}^{'}}}}$ of the state ${\rho _{{A_1}{B_{1}^{'}}}}$ shared by Alice and Bob is
\begin{equation}
{\gamma _{{A_1}\!{B_{1}^{'}}}}\! = \!\left( \!{\begin{array}{*{20}{c}}
{\!{V_1}\!{I_2}}&{\sqrt {T(V_1^2\! -\! 1)}\! {\sigma _z}}\\
\!{\sqrt {T(V_1^2 \!- \!1)} \!{\sigma _z}}&{[T({V_1}\! -\! 1)\! +\! 1 \!+ \!T\varepsilon ']\!{I_2}}
\end{array}} \right)
\end{equation}
where \[T = \frac{{{T_1}}}{2}{g^2},\]
\begin{eqnarray*}
\varepsilon ' = &&1 + \frac{1}{{{T_1}}}\left[ {2 + {T_2}\left( {{\varepsilon _2} - 2} \right) + {T_1}\left( {{\varepsilon _2} - 1} \right)} \right] \nonumber\\
&&+ \frac{1}{{{T_1}}}{\left( {\frac{{\sqrt 2 }}{g}\sqrt {{V_B}}  - \sqrt {{T_2}} \sqrt {{V_B} + 2} } \right)^2}.
\end{eqnarray*}
Here one selects $g = \sqrt {\frac{2}{{{T_2}}}} \sqrt {\frac{{{V_B}}}{{{V_B} + 2}}}$ \cite{li2014continuous} so that the equivalent excess noise $\varepsilon '$ is minimal. So there is:
\begin{equation}
\varepsilon ' = {\varepsilon _1} + \frac{1}{{{T_1}}}\left[ {{T_2}\left( {{\varepsilon _2} - 2} \right) + 2} \right].
\end{equation}
Accordingly, the mutual information between Alice and Bob has the following form:
\begin{equation}
{I_{AB}} = \log_2 \left[ {\frac{{T\left( {V + \chi } \right) + 1}}{{T\left( {1 + \chi } \right) + 1}}} \right],
\end{equation}
where $\chi  = \frac{1}{T} - 1 + \varepsilon '$.

In the equivalent one-way protocol model, the analysis method of the finite-size effect on the parameter estimation is consistent with the CV-QKD protocol. Within this model, Alice and Bob's data respectively satisfies the Gaussian distribution and their data are linked through the following relation:
\begin{equation}
y=tx+z,
\end{equation}
where $t = \sqrt T $ and $z$ follows a centered normal distribution with unknown variance ${\sigma ^2} = 1 + T\epsilon '$. For any value of the modulation variance ${V_A}$, $S(b:E)$ and the variables $t$, ${\sigma ^2}$ are related as follows:
\begin{equation}
{\left. {\frac{{\partial S\left( {b:E} \right)}}{{\partial t}}} \right|_{{\sigma ^2}}} < 0, {\left. {\frac{{\partial S\left( {b:E} \right)}}{{\partial {\sigma ^2}}}} \right|_t} > 0.
\end{equation}
This means that, under $1 - {\epsilon _{PE}}$ probability there is a covariance matrix that minimizes the secret key rate:
\begin{equation}
{\gamma _{{A_1}\!{B_{1}^{'}}}} = \left( {\begin{array}{*{20}{c}}
{\left( {{V_A} + 1} \right){I_2}}&{{t_{\min }}Z{\sigma _z}}\\
{{t_{\min }}Z{\sigma _z}}&{\left( {t_{\min }^2{V_A} + \sigma _{\max }^2} \right){I_2}}
\end{array}} \right),
\end{equation}
where ${V_A}$ is the modulation variance of Alice. And for Gaussian modulation, the parameter $Z = {\left( {{V_A}^2 + 2{V_A}} \right)^{{1 \mathord{\left/
 {\vphantom {1 2}} \right.
 \kern-\nulldelimiterspace} 2}}}$. To analyze the impact of the statistical fluctuation of each parameter on the covariance matrix, the parameters to be estimated are substituted into the above matrix. The covariance matrix is changed into:
\begin{widetext}
\begin{equation}
{\gamma _{{A_1}\!{B_{1}^{'}}}} = \left( {\begin{array}{*{20}{c}}
{\left( {{V_A} + 1} \right){I_2}}&{\frac{{t}_{1}'}{{t}_{2}'}\sqrt {\frac{{{V_B}}}{{{V_B} + 2}}} Z{\sigma _z}}\\
{\frac{{t}_{1}'}{{t}_{2}'}\sqrt {\frac{{{V_B}}}{{{V_B} + 2}}} Z{\sigma _z}}&{\left[ {\frac{{t'_{1}}^{2}}{{t'_{2}}^{2}}\frac{{{V_B}}}{{{V_B} + 2}}{V_A} + 1 + \frac{{{V_B}}}{{{V_B} + 2}}\left( {\frac{{{\sigma '_{1}}^{2} + {\sigma '_{2}}^{2} - 2{t '_{2}}^{2}}}{{t_2^{'2}}}} \right)} \right]{I_2}}
\end{array}} \right).
\end{equation}
\end{widetext}
When the estimated parameters are determined, we can calculate the maximum value of ${S}_{ {\epsilon}_{PE}}(b:E)$.In consequence, the secret key rate of CV-MDI QKD protocol against collective attack with finite-size effect can be calculated according to the Eqs. (\ref{eq:finiteK}) and (\ref{eq:finite-n}).


\section{SIMULATION RESULTS AND DISCUSSION}
\begin{figure}[t]
\centering\includegraphics[width=3.5in,height=2.5in] {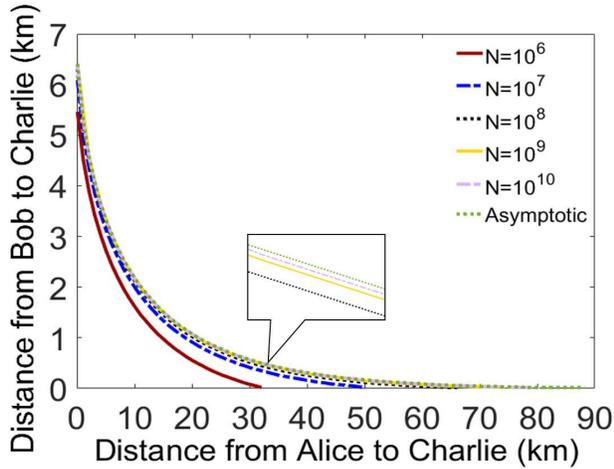}
\caption{\label{fig:epsart}(Color online) Comparison between the maximal transmission distance for the CV-MDI QKD protocol under the finite-size effect. The CV-MDI QKD protocol with ideal reconciliation efficiency $\beta\!=\!1$, where the secret key rate $k$ is positive. The block lengths from left to right curves correspond to $N\!=\!{10^6},{10^7},{10^8},{10^9},{10^{10}}$, and asymptotic regime. Here we use the ideal modulation variance ${V_A}\! =\! {V_B}\! =\! {10^5}$, excess noises ${\varepsilon _1}\! =\! {\varepsilon _2}\! =\! 0.002$ \cite{jouguet2013experimental}.}
\label{fig:LF-eta1}
\end{figure}

In this section, we give and discuss the numerical simulation results of the CV-MDI QKD protocol in ideal detection case with the finite-size effect. See the appendix for the impact of the practical detection on the CV-MDI QKD protocol. To consider the characteristics of CV-MDI QKD protocol, we first perform numerical simulations of the secret key rate in the ideal reconciliation efficiency. In the following figures, the dark red solid lines, dark blue dot-dashed lines, dark black dotted lines, light yellow solid lines, light pink dot-dashed lines, and light green dotted lines correspond to the block lengths of ${10^6},{10^7},{10^8},{10^9},{10^{10}}$, and asymptotic curves, respectively. Figure \ref{fig:LF-eta1} shows the relationship between transmission distance from Alice to Charlie (${L_{AC}}$) and from Bob to Charlie (${L_{BC}}$) of the CV-MDI QKD protocol under finite-size effect. It can be seen that, under different block lengths, when Bob is placed in the untrusted third party (${L_{BC}} = 0$), the asymmetric structure, ${L_{AC}}$ has the farthest transmission distance, more than 85 km. When ${L_{BC}}$ increases, ${L_{AC}}$ decreases rapidly and the sum of the two is also reduced. For $N = {10^{10}}$ (light green dashed line), even if ${L_{AC}}$ is reduced to 0, the farthest ${L_{BC}}$ cannot exceed 7 km.

\begin{figure}[t]
\centering\includegraphics[width=3.5in,height=2.5in]{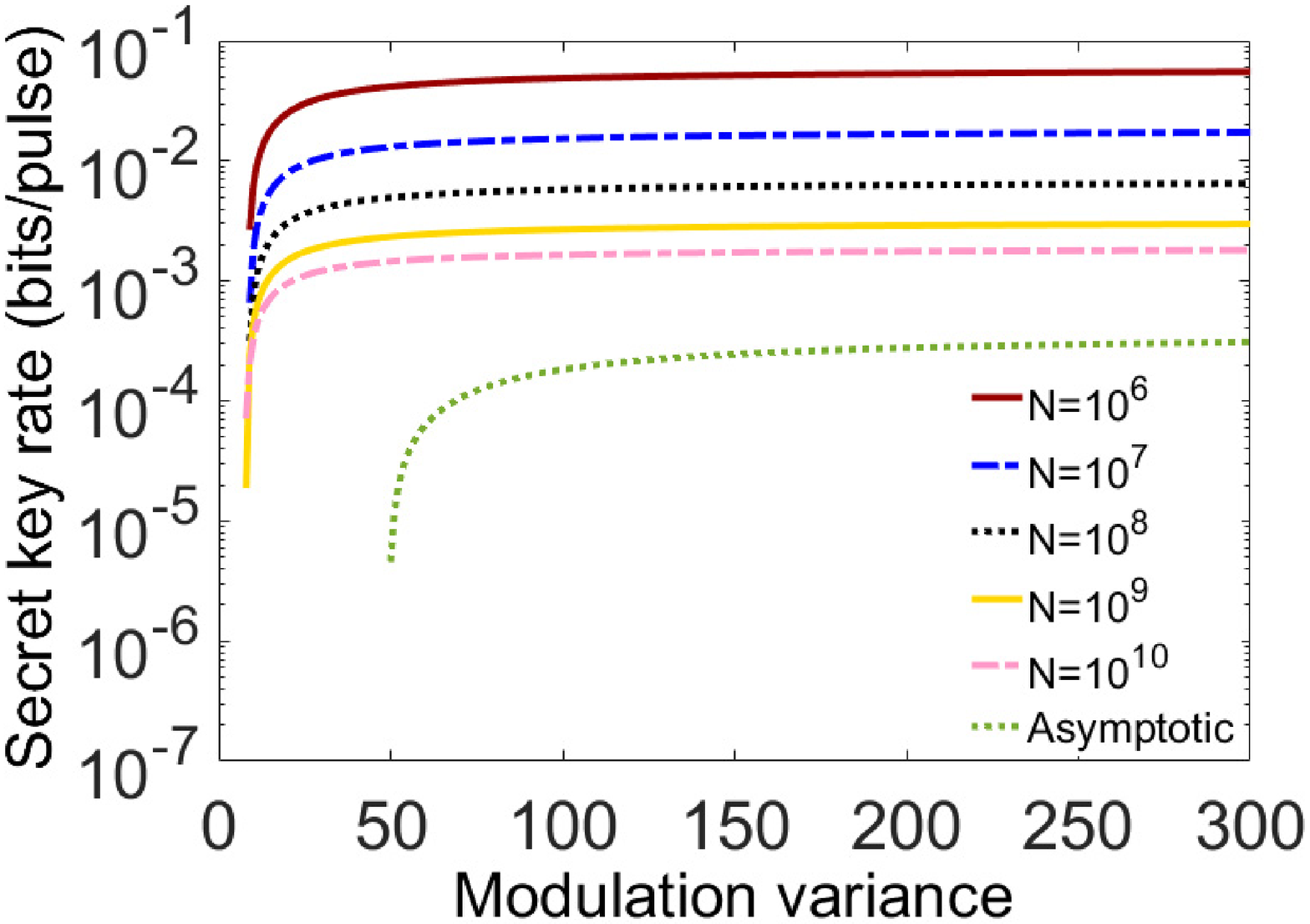}
\caption{\label{fig:epsart}(Color online) Optimal modulation variance for the CV-MDI QKD protocol in the asymmetric case with the finite-size effect. CV-MDI QKD protocol with ideal reconciliation efficiency $\beta  = 1$. From top to bottom, the block length $N$ is equal to $10^6$,$10^7$,$10^8$,$ 10^9$,$10^{10}$, and asymptotic regime. Here we use the excess noises ${\varepsilon _1} = {\varepsilon _2} = 0.002$ \cite{jouguet2013experimental}.}
\label{fig:VF-eta1}
\end{figure}

For the sake of optimizing the performance of the protocol, we also need to consider the optimal modulation variance of the protocol under different block lengths. Figure \ref{fig:VF-eta1} shows the relationship between the modulation variance and the secret key rate under the ideal reconciliation efficiency of the CV-MDI QKD protocol with the finite-size effect. We find that with the improvement of the modulation variance, the secret key rate gradually converges under the ideal reconciliation efficiency ($\beta  = 1$). The optimal modulation variance of the CV-MDI QKD protocol with the finite-size effect in the asymmetric case tends to be infinite. We choose ${V_A} = {V_B} = {10^5}$ to see the performance of the protocol in the ideal modulation.

\begin{figure}[t]
\centering\includegraphics[width=3.5in,height=2.5in]{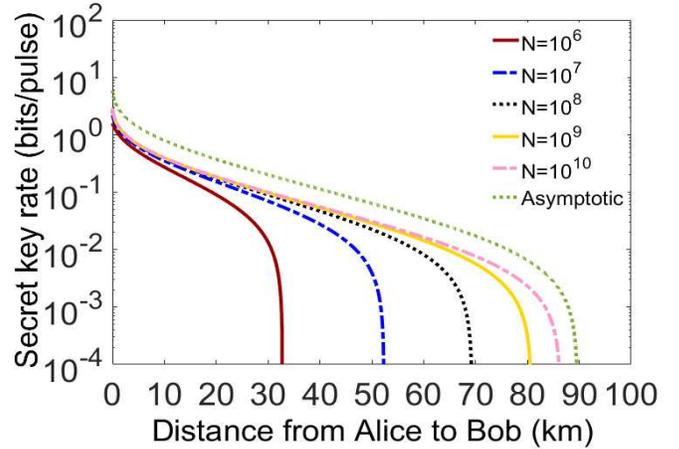}
\caption{\label{fig:epsart}(Color online) Secret key rate for the CV-MDI QKD protocol in the asymmetric case with the finite-size effect. The CV-MDI QKD protocol with ideal reconciliation efficiency $\beta  = 1$. The block lengths from left to right curves correspond to $N = {10^6},{10^7},{10^8},{10^9},{10^{10}}$, and asymptotic regime. Here we use the ideal modulation variance ${V_A} = {V_B} = {10^5}$, excess noises ${\varepsilon _1} = {\varepsilon _2} = 0.002$ \cite{jouguet2013experimental}.}
\label{fig:F-eta1}
\end{figure}

\begin{figure}[htbp]
\centering\includegraphics[width=3.5in,,height=2.5in]{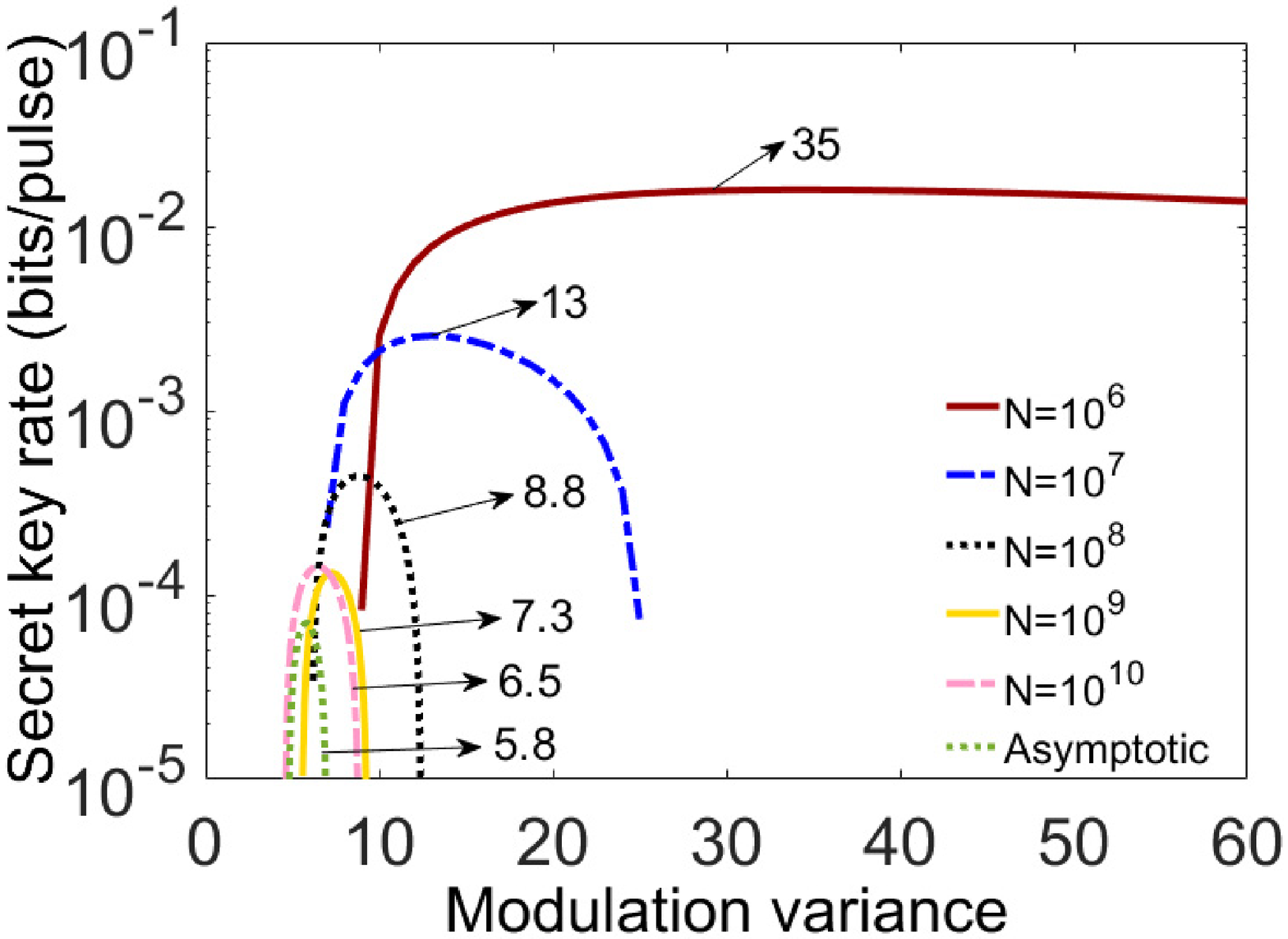}
\caption{\label{fig:epsart}(Color online) Optimal modulation variance for the CV-MDI QKD protocol in asymmetric case with the finite-size effect. The CV-MDI QKD protocol with imperfect reconciliation efficiency $\beta  = 96.9\%$. From top to bottom, the block length $N$ is equal to $10^6$,$10^7$,$10^8$,$ 10^9$,$10^{10}$, and asymptotic regime. Here we use the excess noises ${\varepsilon _1} = {\varepsilon _2} = 0.002$ \cite{jouguet2013experimental}.}
\label{fig:VF-eta969}
\end{figure}

\begin{figure}[htbp]
\centering\includegraphics[width=3.5in,,height=2.5in]{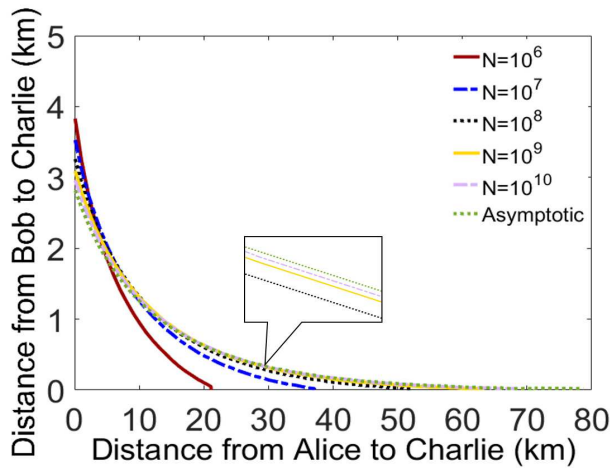}
\caption{\label{fig:epsart}(Color online) Comparison among the maximal transmission distance for the CV-MDI QKD protocol under finite-size effect. The CV-MDI QKD protocol with imperfect reconciliation efficiency $\beta  = 96.9\%$, within which the secret key rate $k$ is positive. The block lengths from left to right curves correspond to $N = {10^6},{10^7},{10^8},{10^9},{10^{10}}$, and asymptotic regime. Here we use the optimal modulation variance, excess noises ${\varepsilon _1} = {\varepsilon _2} = 0.002$ \cite{jouguet2013experimental}.}
\label{fig:LF-eta969}
\end{figure}

Figure \ref{fig:F-eta1} displays the impact of the finite-size effect on the CV-MDI QKD protocol under the asymmetric case and ideal modulation variance. As shown in the figure, when the data length is $N = {10^6}$ (dark red solid line), the security transmission distance is about 32 km. When the data length is $N = {10^{10}}$ (light pink dot-dashed line), the security transmission distance of the protocol is up to 86 km, and the longer the data length, the closer the security transmission distance is to asymptotic regime.

Noteworthily, in Eq. (\ref{eq:finiteK}), the reconciliation efficiency $\beta $ can be regarded as the ratio between the actual extracted mutual information and the ideal extracted mutual information \cite{leverrier2008multidimensional}. After data reverse reconciliation, the length of the key shared between Alice and Bob becomes $n\beta I\left( {x:y} \right)$. Therefore, under the finite-size effect, imperfect reconciliation efficiency will also affect the security transmission distance of the CV-MDI QKD protocol. Figure \ref{fig:VF-eta969} shows the optimal modulation variance of the CV-MDI QKD protocol with imperfect reconciliation efficiency $\beta {\rm{ = 96}}{\rm{.9\% }}$ \cite{jouguet2011long,Wang2017Efficient} in different block lengths. The numbers at the end of the arrow are the values of the optimal modulation variance. The simulation results show that the finite-size effect has an influence on the optimal modulation variance under the imperfect reconciliation efficiency. With the increase of block length, the optimal modulation variance is decreasing.

\begin{figure}[t]
\centering\includegraphics[width=3.5in,height=2.5in]{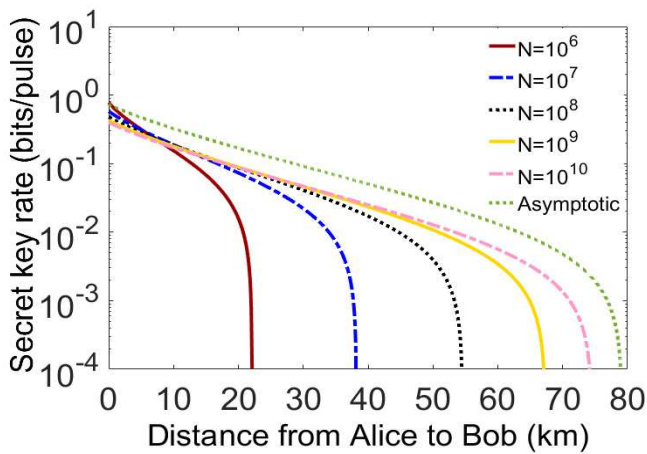}
\caption{\label{fig:epsart}(Color online) Secret key rate for CV-MDI QKD protocol in asymmetric case with finite-size effect. CV-MDI QKD protocol with imperfect reconciliation efficiency $\beta  = 96.9\%$. The block lengths from left to right curves correspond to $N = {10^6},{10^7},{10^8},{10^9},{10^{10}}$ and asymptotic regime. Here we use the optimal modulation variance, excess noises ${\varepsilon _1} = {\varepsilon _2} = 0.002$ \cite{jouguet2013experimental}.}
\label{fig:F-eta969}
\end{figure}

Similarly, in the imperfect reconciliation efficiency, the asymmetric case of the CV-MDI QKD protocol has the farthest security transmission distance, as shown in Fig. \ref{fig:LF-eta969}. And the selection of the optimal modulation variance increases the final security transmission distance. In Fig. \ref{fig:F-eta969}, we display the relationship between secret key rate and transmission distance of the CV-MDI QKD protocol in the asymmetric case with the finite-size effect under imperfect reconciliation efficiency and optimal modulation variance. The simulation results show that the optimal modulation variance improves the security transmission distance while increasing the secret key rate. And when the data length is ${10^6}$ (dark red solid line), the security transmission distance is about 23 km.When the data length is ${10^{10}}$ (light pink dot-dashed line), the protocol has a far greater security transmission distance, close to 75 km in the case of imperfect reconciliation efficiency ($\beta  = 96.9\%$).

\section{\label{sec:additional}CONCLUSION}
In this paper, we roughly describe the CV-MDI QKD protocol and propose a finite-size analysis of the CV-MDI QKD protocol under collective attack. This provides a bridge for the theoretical and practicality of the CV-MDI QKD protocol. By using the continuous-variable protocol under the finite-size scenario of the secret key rate formula for numerical simulation, we see that the secret key rate and the security transmission distance are affected when considering the finite-size effect. The results demonstrate that the finite-size effect also influences the optimal modulation variance. With the increase of the block length, the optimal modulation variance is decreasing. And at various block lengths between ${10^6}$ and ${10^{10}}$, when Bob is placed in an untrusted third party; that is, under the asymmetric case, the CV-MDI QKD protocol has the farthest security transmission distance. The CV-MDI QKD protocol of the asymmetric structure with the finite-size effect can safely transmit about 86 km under the ideal reconciliation efficiency and optimal modulation variance conditions at ${10^{10}}$ block size . When the reconciliation efficiency is 96.9\%, under the above conditions, the maximum transmission distance of the protocol is about 75 km.

It can be seen that, as with the CV-QKD one-way protocol, in the CV-MDI QKD protocol, the fewer signals exchanged, the more obvious is the finite-size effect, the faster the secret key rate and security transmission distance drop. The overall results show that the practical implementation of CV-MDI QKD should not neglect the influence of the finite-size effect.

Note that we mainly focus on the impact of the finite-size effect on the CV-MDI QKD protocol under arbitrary collective attack. Recently, the composable security of CV-MDI QKD protocol against coherent attacks in a practical finite-size scenario has been proved rigorously \cite{Lupo2017Composable,zhang2017}.

Note added. An independent work \cite{Papanastasiou2017finite} has been posted on arXiv. This work also studies the impact of the finite-size effect of the CV-MDI QKD protocol.

\begin{acknowledgments}
This work was supported in part by the National Basic Research Program of China (973 Program) under Grant 2014CB340102, and in part by the National Natural Science Foundation under Grants 61531003 and 61427813.
\end{acknowledgments}

\begin{appendices}
\section*{APPENDIX: Discussion on practical detection}

We now consider the finite-size effect related to the detection setup. Since the third party Charlie uses two homodyne detectors to measure the quantum state for the Bell measurement in the practical system, the impact of the practical detectors on the CV-MDI QKD protocol should be considered. Although the security of the CV-MDI QKD protocol is not limited by the detector, its performance is constrained by the performance of the detector. To research the effect of the practical detection on the protocol in the case of the finite-size effect, we mainly consider the ideal coordination efficiency ($\beta {\rm{ = }}1$) situation. Now, the EB version of the CV-MDI QKD protocol in a practical detection scheme is shown in Fig. \ref{fig:EB}. An imperfect homodyne detection is represented by a beam splitter and thermal noise, wherein the transmittance of the beam splitter is the detection efficiency $\eta $, the relationship between the variance of the thermal state $v$ and detector electronic noise variance ${v_{el}}$ is $v = 1 + {v_{el}}/(1 - \eta )$. For the CV-MDI QKD protocol considering the practical detector, it is also necessary to estimate the covariance matrix between Alice and Bob. The estimated parameters are the variance of Alice, Bob, Charlie, Alice and Charlie's covariance, and Bob and Charlie's covariance:
\begin{equation}
\left\langle {x_1^2} \right\rangle  = {V_1} - 1 = {V_A},\left\langle {x_2^2} \right\rangle  = {V_2} - 1 = {V_B},
\end{equation}
\begin{eqnarray}
\left\langle {y_1^2} \right\rangle=&& \left\langle {y_2^2} \right\rangle =  \eta \left[ {\frac{1}{2}{T_1}({V_1} + {\chi _1}) + \frac{1}{2}{T_2}({V_2} + {\chi _2})} \right]\nonumber\\
\nonumber\\&&
+ 1 - \eta  + {v_{el}}\nonumber\\
\nonumber\\
=&&
\frac{1}{2}\eta ({T_1}{V_A} + {T_2}{V_B}) + \frac{1}{2}\eta ({T_1}{\varepsilon _1} + {T_2}{\varepsilon _2})\nonumber\\
\nonumber\\&&
+ 1 + {v_{el}},
\end{eqnarray}
\begin{eqnarray}
\left\langle {{x_1}{y_1}} \right\rangle  = \sqrt \eta  \sqrt {\frac{{{T_1}}}{2}} ({V_1} - 1) = \sqrt {\frac{{\eta {T_1}}}{2}} {V_A},\nonumber\\
\left\langle {{x_2}{y_2}} \right\rangle  = \sqrt \eta  \sqrt {\frac{{{T_2}}}{2}} ({V_2} - 1) = \sqrt {\frac{{\eta {T_2}}}{2}} {V_B},
\end{eqnarray}
~\\
\begin{eqnarray}
\left\langle {{y_1}{y_2}} \right\rangle  =&& \eta \left[ {\frac{1}{2}{T_1}({V_1} + {\chi _1}) - \frac{1}{2}{T_2}({V_2} + {\chi _2})} \right] \nonumber\\
=&& \frac{1}{2}\eta ({T_1}{V_A} - {T_2}{V_B}) + \frac{1}{2}\eta ({T_1}{\varepsilon _1} - {T_2}{\varepsilon _2}).
\end{eqnarray}
~\\
\begin{figure}[t]
\centering\includegraphics[width=3.5in,height=2.5in]{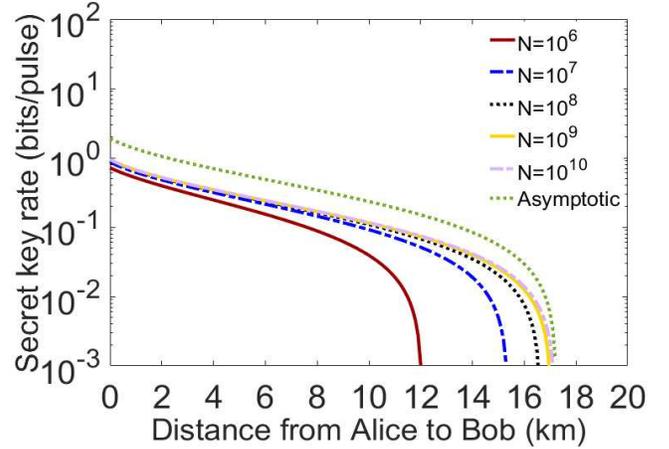}
\caption{\label{fig:epsart}(Color online) Secret key rate for CV-MDI QKD protocol in asymmetric case with finite-size effect. CV-MDI QKD protocol with perfect reconciliation efficiency $\beta  = 1$ and imperfect homodyne detectors $\eta  = 96\%,{\upsilon _{el}} = 0.015$. The block lengths from left to right curves correspond to $N = {10^6},{10^7},{10^8},{10^9},{10^{10}}$ and asymptotic regime.  Here we use the ideal modulation variance ${V_A} = {V_B} = {10^5}$, excess noises ${\varepsilon _1} = {\varepsilon _2} = 0.002$ \cite{jouguet2013experimental}.}
\label{fig:F-praceta96}
\end{figure}

The parameter estimations below are calculated in the same way as the calculation without considering the detections, but the statistical fluctuations of $\eta $ and ${v_{el}}$ are taken into account. Correspondingly, the equivalent excess noise is written as\cite{li2014continuous}:
\begin{equation}
\varepsilon '{\rm{ = }}1{\rm{ + }}\frac{{{T_1}{\chi _1} + {T_2}{\chi _2} - {T_2}}}{{{T_1}}} + \frac{{2{\chi _3}}}{{{T_1}}},
\end{equation}
where ${\chi _3} = \frac{{1 - \eta }}{\eta } + \frac{{{v_{el}}}}{\eta }$, and the optimized displacement operation amplification coefficient is $g = \sqrt {\frac{2}{{\eta {T_2}}}} \sqrt {\frac{{{V_B}}}{{{V_B} + 2}}} $.

Figure \ref{fig:F-praceta96} gives the impact of the finite-size effect on the CV-MDI QKD protocol considering the practical detection. Compared with the results of Fig. \ref{fig:F-eta1}, we found that imperfect detectors have a significant effect on the secret key rate and the security transmission distance. For instance, when the block length is ${10^{10}}$, the security transmission distance is the farthest, only 17km. Conceivably, the results will be even more pessimistic in terms of practical reconciliation efficiency.
\end{appendices}
%

\end{document}